\documentstyle[preprint,aps,epsf,tighten]{revtex}

\newcommand{\slashed}[1]{\rlap{$#1$}/}
\newcommand{\slashp}{\mbox{$\not \hspace*{-1.10mm} p$}}

\newcommand{\GeV}{\mbox{\rm GeV}}

\newcommand{\tr}{\mbox{\rm tr}}
\newcommand{\lsim}[1]{
\setlength{\unitlength}{12pt}
\begin{picture}(1.4,1.)
\put(.7,-0.3){\makebox(0.0,1.)[t]{$<$}}
\put(.7,-0.3){\makebox(0.0,1.)[b]{$\sim$}}
\end{picture}#1}

\begin{document}


\preprint{ZTF-98/04}


\title{
$\gamma^\star\gamma\to\pi^0$ transition and asymptotics of 
$\gamma^\star\gamma$ and $\gamma^\star\gamma^\star$ \\
transitions of other unflavored pseudoscalar mesons
}

\author{Dalibor Kekez}
\address{\footnotesize Rudjer Bo\v{s}kovi\'{c} Institute,
         P.O.B. 1016, 10001 Zagreb, Croatia}

\author{Dubravko Klabu\v{c}ar}
\address{\footnotesize Department of Physics, Faculty of Science, \\
        Zagreb University, P.O.B. 162, 10001 Zagreb, Croatia}

\maketitle

\begin{abstract}
\noindent 
For the spacelike momenta $k$ of the virtual photon $\gamma^\star$,
the $\pi^0(p)\gamma^\star(k)\gamma(k^{\prime})$ transition form factor 
is considered in the coupled Schwinger-Dyson and Bethe-Salpeter
approach in conjunction with the generalized impulse approximation
using the dressed quark--photon--quark vertices of the Ball-Chiu
and Curtis-Pennington type.
These form factors are compared with the ones predicted by the vector 
meson dominance, operator product expansion, QCD sum rules, and the
perturbative QCD for the large spacelike transferred momenta $k$.
The most important qualitative feature of the asymptotic
behavior, namely the $1/k^2$ dependence, is in our approach 
obtained in the model-independent way. Again model-independently,
our approach reproduces also the Adler-Bell-Jackiw anomaly result
for the limit of both photons being real. 
For the case of one highly virtual photon, we find in the closed
form the asymptotic expression which can be easily generalized both to
the case of other unflavored pseudoscalar mesons
$P^0 = \pi^0, \eta_8, \eta_0, \eta_c, \eta_b$, and to the case
of arbitrary virtuality of the other photon.
Implications thereof for certain important theoretical
and experimental applications are pointed out.

\end{abstract}

\vspace{3mm}

vertices; Chiral symmetry breaking; 
QCD; QED axial anomaly; Two--photon processes

\vskip 1.5cm

\newpage

\noindent {\bf 1.} The interest in the form factor 
$T_{\pi^0}(-Q^2,0)$ for the transition
$\gamma^\star(k)\gamma(k^{\prime})\to\pi^0(p)$ (where
$k^{2} = - Q^2 \neq 0$ is the momentum-squared of the
spacelike off-shell photon $\gamma^\star$), 
has again been growing lately for both experimental
(the new CLEO data \cite{gronberg98} and
plans for new TJNAF measurements \cite{gagasCEBAF}) and 
theoretical reasons -- such as its relevance for the
hadronic light--by--light scattering contribution to
muon $g-2$ \cite{H+Kinoshita98}, which is in turn again 
relevant for the experiment E821 at BNL 
\cite{Miller+al97}.
Nevertheless, this transition form factor still represents
a more demanding theoretical challenge than usually thought,
as the perturbative QCD (pQCD) may be not sufficient
even at the highest of the presently accessible momenta
($Q^2 \lsim 10$ GeV$^2$) -- {\it e.g.}, see Refs. \cite{Radyushkin+Rusk3}.
These references therefore indicate desirability of having
direct calculations of $T_{\pi^0}(-Q^2,0)$ without any 
{\it a priori} assumptions about the shape of the pion
distribution amplitude $\varphi_\pi$. 
Even more important for the present work is that
the pQCD approaches, oriented mostly at reproducing the large $Q^2$
behavior, have problems at low $Q^2$. This was most recently 
stressed 
in Ref. \cite{M+Rady97} discussing
the status of QCD--based theoretical predictions for
this process.
See especially its Sec. IV for clarifications how such 
approaches \cite{JakobKrollRaulfs96+Cao+al96} can fail to reproduce 
the $Q^2=0$ value (\ref{AnomAmpl}) corresponding to the Abelian 
axial {\it alias} Adler-Bell-Jackiw (ABJ) anomaly 
which explains the 
$\pi^0$ decay into two real photons ($k^{2} = k^{\prime 2} = 0$). 

On the other hand, the approaches (such as the present one)
which rely on the chirally 
well--behaved Schwinger--Dyson (SD) and  Bethe--Salpeter (BS) 
equations{\footnote{See, for example, Ref. \cite{Miransky}, and the 
recent reviews \cite{nucl-th9807026+RW}, or \cite{CahillFizika}.} 
for the light pseudoscalar mesons ($\pi, K, \eta, ...$) while 
taking care that the Ward-Takahashi identities (WTI) of QED are 
respected, satisfy in this respect even what Ref.  
\cite{M+Rady97} calls the ``maximalist" requirement -- namely 
that the fundamental axial anomaly result
\begin{equation}
 T_{\pi^0}(0,0) =\frac{1}{4\pi^2 f_\pi} \, 
\label{AnomAmpl}
\end{equation}
should be satisfied automatically in the chiral (and soft) limit. 
(This is clearly superior to imposing it as an external condition, 
which can by analogy be termed the ``minimalist" attitude.) 
In the  present approach \cite{KeKl1,KlKe2,KeBiKl98}, the Abelian 
axial anomaly amplitude (\ref{AnomAmpl}) is obtained without imposing 
any requirements or constraints on the solutions for the $\pi^0$ wave 
function or the solutions for the quark propagators ({\it e.g.}, see
\cite{bando94,Roberts,nucl-th9807026+RW}). 
Since the anomaly -- and thus also $T_{\pi^0}(0,0)$ 
of Eq. (\ref{AnomAmpl}) -- must not depend on the 
internal structure of the pion, this is exactly as it should be.
What must then be explored in the coupled SD-BS approach is if the 
large $Q^2$ behavior can be satisfactorily understood in that approach. 
In particular, the comparison with the new CLEO data \cite{gronberg98} 
-- at $Q^2$ up to 8 GeV$^2$ -- must be made, since the results of the 
SD-BS approach \cite{KeKl1} (as well as the results of the earlier and 
closely related Ansatz approach of \cite{Frank+al}) have so far been 
compared only with the older CELLO data \cite{behrend91} at $Q^2$ not 
exceeding some 2.5 GeV$^2$. In this letter we will show how the SD-BS 
approach to modeling QCD provides a good description for both low and 
high values of $Q^2$. In other words, it shows
how one can in fact {\it obtain} something similar to the 
Brodsky--Lepage (BL) interpolation formula 
$T_{\pi^0}(-Q^2,0)=(1/4\pi^2 f_\pi) \, / \, (1 + Q^2/8\pi^2 f_\pi^2)$
proposed \cite{BrodskyLepage} as a desirable behavior for the 
transition form factor
because it reduces to the ABJ anomaly amplitude (\ref{AnomAmpl}) 
at $Q^2=0$, while agreeing with the following type of leading behavior
for large $Q^2$:
\begin{equation}
 T_{\pi^0}(-Q^2,0) = {\cal J} \, \frac{f_\pi}{Q^2} \,
\qquad ({\cal J} = {constant} \,\, {\rm for \,\, large \,\,} Q^2),
\label{largeQ2}
\end{equation}
which is favored both experimentally \cite{gronberg98} 
and theoretically ({\it e.g.}, see the QCD-based predictions in Refs. 
\cite{BrodskyLepage,ChernyakZhit,manohar90,Radyushkin+Rusk3,JakobKrollRaulfs96+Cao+al96,M+Rady97}, and references therein.) 

\noindent {\bf 2.}
In the coupled SD-BS approach, the BS equation for the pion bound-state 
$q\bar q$ vertex $\Gamma_{\pi^0}(q,p)$ employs the dressed quark propagator 
$S(k) = [ A(k^2)\slashed{k} - B(k^2) ]^{-1}$,
obtained by solving its SD equation. Solving the SD and BS equations 
in a consistent approximation is crucial ({\it e.g.}, see Refs. 
\cite{jain91+munczek92,jain93b,Bender+al96,MarisRoberts97PRC56})
for obtaining $q\bar q$ bound states which are, in the case of light 
pseudoscalar mesons, simultaneously also the (pseudo-)Goldstone bosons of 
dynamical chiral symmetry breaking (D$\chi$SB).

Following Jain and Munczek \cite{jain91+munczek92,jain93b}, we 
adopt the ladder-type approximation
sometimes called the {\it improved} \cite{Miransky} or {\it generalized} 
\cite{Roberts} ladder approximation (employing bare quark--gluon--quark 
vertices but dressed propagators). For the gluon propagator we use an effective,
({\it partially}) modeled one in Landau-gauge \cite{jain91+munczek92,jain93b}, 
given by
$G(-l^2)( g^{\mu\nu} - {l^\mu l^\nu}/{l^2} )~.$
(This Ansatz is often called the 
``Abelian approximation" \cite{MarisRoberts97PRC56}.)
What is essential is that the effective propagator function $G$ is the
sum of the perturbative contribution $G_{UV}$ and the nonperturbative
contribution $G_{IR}$:
$G(Q^2) = G_{UV}(Q^2) + G_{IR}(Q^2)~,\;\;(Q^2 = -l^2)~.$
The perturbative part $G_{UV}$ is required to reproduce correctly
the ultraviolet (UV) asymptotic behavior that unambiguously
follows from QCD in its high--energy regime.
Therefore, this part must be given
-- up to the factor $1/Q^2$ --
by the running coupling constant $\alpha_s(Q^2)$ which is
well-known from perturbative QCD, so that $G_{UV}$ is 
{\it not} modeled.
As in Refs. \cite{KeKl1,KlKe2,KeBiKl98}, we follow 
Refs. \cite{jain91+munczek92,jain93b} and employ the 
two--loop asymptotic expression for $\alpha_s(Q^2)$.
For the modeled, IR part of the gluon propagator, we 
adopt from Ref.~\cite{jain93b}
$G_{IR}(Q^2)=(16\pi^2/3) \,a\,Q^2 e^{-\mu Q^2}$, 
with their parameters
$a=(0.387\,\GeV)^{-4}$ and $\mu=(0.510\,\GeV)^{-2}$.

More details on our calculational procedures can be found in our 
Refs. \cite{KeKl1,KlKe2,KeBiKl98}.
We essentially reproduce Jain and Munczek's \cite{jain93b} solutions 
for the dressed propagators $S(q)$, i.e., the functions $A(q^2)$ and 
$B(q^2)$, as well as the solutions for the four functions comprising 
the pion bound-state vertex $\Gamma_{\pi^0}$. 
Actually, Ref. \cite{jain93b} employs the BS amplitude 
$\chi_{\pi^0}(q,p)\equiv S(q+{p}/{2})\Gamma_{\pi^0}(q,p)S(q-{p}/{2})$,
which is completely equivalent.

\noindent {\bf 3.} 
We assume that the $\pi^0\gamma^\star\gamma$ transition 
proceeds through the triangle graph (Fig. 1), and 
that we calculate the pertinent amplitude 
$T_{\pi^0}^{\mu\nu}(k,k^\prime) = \varepsilon^{\alpha\beta\mu\nu}
 k_\alpha k^\prime_\beta T_{\pi^0}(k^2,k^{\prime 2})~$
as in Refs. \cite{KeKl1,KlKe2,KeBiKl98},
using the framework advocated by (for example)
\cite{bando94,Roberts,Frank+al,Burden+al,AlkRob96}
in the context of electromagnetic interactions of BS bound states,
and often called the generalized impulse approximation (GIA) -
{\it e.g.}, by \cite{Frank+al,Burden+al}.
To evaluate the triangle graph, we therefore use the
{\it dressed} quark propagator $S(q)$
and the pseudoscalar BS bound--state vertex $\Gamma_P(q,p)$.
Another ingredient, crucial for GIA's ability to
reproduce the correct Abelian anomaly result, is
employing an appropriately dressed
{\it electromagnetic} vertex $\Gamma^\mu(q^\prime,q)$,
which satisfies the vector Ward--Takahashi identity (WTI)
$(q^\prime-q)_\mu \Gamma^\mu(q^\prime,q)=S^{-1}(q^\prime)-S^{-1}(q)~.$
Namely, assuming that photons
couple to quarks through the bare vertex $\gamma^\mu$
would be inconsistent with
our quark propagator $S(q)$, which, dynamically dressed through
its SD-equation, contains the momentum-dependent
functions $A(q^2)$ and $B(q^2)$.
The bare vertex $\gamma^\mu$ obviously violates the vector WTI, implying 
the nonconservation of the electromagnetic vector current and charge.
Solving the pertinent SD equation for the dressed quark--photon--quark 
($qq\gamma$) vertex $\Gamma^\mu$ is a difficult and still unsolved problem,
and using the realistic Ans\"{a}tze for $\Gamma^\mu$ still remains the only 
practical way to satisfy the WTI. 
The simplest solution of the vector WTI is the Ball--Chiu (BC) ~\cite{BC} 
vertex 
        \begin{eqnarray}
        \Gamma^\mu_{BC}(q^\prime,q) =
        A_{\bf +}(q^{\prime 2},q^2)
       \frac{\gamma^\mu}{\textstyle 2}
        + \frac{\textstyle (q^\prime+q)^\mu }
               {\textstyle (q^{\prime 2} - q^2) }
        \{A_{\bf -}(q^{\prime 2},q^2)
        \frac{\textstyle (\slashed{q}^\prime + \slashed{q}) }{\textstyle 2}
         - B_{\bf -}(q^{\prime 2},q^2) \}~,
        \label{BC-vertex}
        \end{eqnarray}
where
$H_{\bf \pm}(q^{\prime 2},q^2)\equiv [H(q^{\prime 2})\pm H(q^2)]$,
for $H = A$ or $B$.
This particular solution of the vector WTI reduces to the bare 
vertex in the free-field limit as must be in perturbation theory, has 
the same transformation properties under Lorentz transformations and 
charge conjugation as the bare vertex, and has no kinematic singularities. 
Note that it does not introduce any new parameters as it is completely
determined by the dressed quark propagator $S(q)$. 
In phenomenological calculations in the SD-BS approach,
this minimal WTI-satisfying Ansatz (\ref{BC-vertex}) is 
still the most widely used $qq\gamma$ vertex
({\it e.g.}, Refs. \cite{Frank+al,Burden+al,Roberts,AlkRob96}).
A general WTI-satisfying vertex can be written \cite{BC} as 
$\Gamma^\mu = \Gamma^\mu_{BC} + \Delta \Gamma^\mu$,
where the addition $\Delta \Gamma^\mu$ does not contribute
to the WTI, since it is transverse, 
$(q^\prime-q)_\mu \Delta \Gamma^\mu(q^\prime,q) = 0$.
That is, $\Delta \Gamma^\mu(q^\prime,q)$ entirely lies
in the hyperplane spanned by the eight vectors $T_i^\mu(q^\prime,q)$
transverse to the photon momentum $k=q^\prime-q$.
Curtis and Pennington (CP) \cite{CP90} advocated a transverse Ansatz for
$\Delta \Gamma^\mu(q^\prime,q)$ exclusively along the  
basis vector usually denoted $T_6^\mu(q^\prime,q)$:
\begin{equation}
\Delta \Gamma^\mu(q^\prime,q) = T_6^\mu(q^\prime,q) 
\frac{A_{\bf -}(q^{\prime 2},q^2)}{2 d(q^\prime,q)} \, ; 
\qquad T_6^\mu(q^\prime,q) \equiv \gamma^\mu(q^{\prime 2} - q^2)
- (q^\prime + q)^\mu ( \slashed{q}^\prime - \slashed{q} ) \, .
\label{defDelta}
\end{equation}
Then, the coefficient multiplying $T_6^\mu(q^\prime,q)$ can be suitably 
chosen to ensure multiplicative renormalizability in the context of 
solving fermion SD equations beyond the ladder approximation in QED$_4$
\cite{CP90}. To this end, $d(q^\prime,q)$ should be a symmetric, 
singularity free function of $q^\prime$ and $q$, with the limiting 
behavior $\lim_{q^2>>{q^\prime}^2} d(q^\prime,q) = q^2$; for example, 
\begin{equation}
d_{\pm}(q^\prime,q) =
        \frac{1}{q^{\prime 2}+q^2}
        \left\{
                (q^{\prime 2} \pm q^2)^2
                +
                \left[ M^2({q^\prime}^2) + M^2(q^2) \right]^2
        \right\}~,
\label{defdfunct}
\end{equation}
where $M(q^2) \equiv B(q^2)/A(q^2)$ is the D$\chi$SB-generated
dynamical mass function, which in our case has the large-$q^2$ 
dependence \cite{KeBiKl98} in agreement with pQCD.

The choice $d = d_-$ corresponds to the CP vertex Ansatz
$\Gamma^\mu_{CP}$ suggested in Ref. \cite{CP90}. 
We will use it in analytic calculations of
$T(-Q^2,0)$, which are possible for $Q^2=0$ and $Q^2\to \infty$.
However, in the numerical calculations, which are 
necessary for finite values of $Q^2\neq 0$, we prefer to use the 
{\it modified} CP (mCP) vertex, $\Gamma^\mu_{mCP}$, resulting from 
the choice $d = d_+$ in Eq. (\ref{defDelta}).
Namely, although the original CP denominator function 
$d_-(q^{\prime},q)$ never vanishes {\it strictly},
the dynamical masses it contains become negligible 
with growing momenta, so that $d_-(q^{\prime},q)$
can become arbitrarily small for the large spacelike 
quark loop momenta $q,q^{\prime}$ if simultaneously 
$q^{\prime 2} - q^2$ is small, causing numerical problems.
On the other hand, the numerical calculations of $T(-Q^2,0)$
employing our standard computational methods \cite{KlKe2,KeBiKl98}
and using the mCP vertices have no such problems
and they are as reliable as those using the minimal, BC vertices.
In contrast to the BC one, the mCP vertex is consistent with 
multiplicative renormalizability by the same token as the CP one.
It should also be noted that $\Gamma^\mu_{mCP}$ is essentially 
equal{\footnote{Up to the inclusion of the mass functions $M(q^2)$, 
and for the simplest choice of their \cite{Cudell+al95} $\eta$-function: 
$\eta(q^{\prime},q) \equiv 1$, {\it i.e.}, their choice $n=0$ for 
the exponent in $\eta$.}} to the high-$q^2$ or ${q^\prime}^2$ leading 
part of the vertex of Cudell {\it et al.} \cite{Cudell+al95} 
[see their Eq. (3.14), and comments below it]. 

In the present context,
the important {\it qualitative} difference between the BC vertex on
one side, and the CP as well as the modified,  mCP vertex 
on the other side, will be that $\Gamma^\mu_{BC}(q^\prime,q)\to\gamma^\mu$
when {\it both} $q^{\prime 2},q^2\to \pm \infty$, whereas 
$\Gamma^\mu_{CP}(q^\prime,q) \to \gamma^\mu$ and 
$\Gamma^\mu_{mCP}(q^\prime,q) \to \gamma^\mu$ 
as soon as {\it one} of the squared momenta tends to infinity.

The correct ABJ anomaly result (\ref{AnomAmpl}) cannot be obtained 
analytically in the chiral limit unless such a $qq\gamma$ vertex 
satisfying the WTI is used \cite{bando94,Roberts}.   
We have checked by the explicit GIA calculation that the 
$\pi^0\to\gamma\gamma$ amplitude (\ref{AnomAmpl}) is 
analytically reproduced employing the CP and mCP vertices, 
in the same way as when the BC vertices were employed in earlier
applications, {\it e.g.} \cite{Roberts,Frank+al,KeKl1,KlKe2}.

In the case of $\pi^0$, GIA yields ({\em e.g.}, see Eq. (24) in 
Ref. \cite{KlKe2}) the amplitude $T_{\pi^0}^{\mu\nu}(k,k^\prime)$:
\begin{displaymath}
        T_{\pi^0}^{\mu\nu}(k,k^\prime)
        =
        -
        N_c \,
        \frac{1}{3\sqrt{2}}
        \int\frac{d^4q}{(2\pi)^4} \mbox{\rm tr} \{
        \Gamma^\mu(q-\frac{p}{2},k+q-\frac{p}{2})
        S(k+q-\frac{p}{2})
\end{displaymath}
\begin{equation}
          \qquad
        \times
        \Gamma^\nu(k+q-\frac{p}{2},q+\frac{p}{2})
        \chi(q,p) \}
        +
        (k\leftrightarrow k^\prime,\mu\leftrightarrow\nu).
\label{Tmunu(2)}
\end{equation}
Here, $\chi$ is the BS amplitude of both $u\bar u$ and $d\bar d$
pseudoscalar bound states: $\chi \equiv \chi_{u\bar u} = \chi_{d\bar d}$
thanks to the isospin symmetry assumed here. This symmetry likewise
enables us to continue suppressing flavor labels also on the quark 
propagators $S$ and $qq\gamma$ vertices $\Gamma^\mu$.
We follow the conventions of Ref. \cite{KlKe2}, including those 
for the flavor factors and flavor matrices $\lambda^a$. Then, 
$\chi_{\pi^0}(q,p) \equiv \chi(q,p) \, \lambda^3/\sqrt{2}$,
so that the prefactor $1/3\sqrt{2}$ in Eq. (\ref{Tmunu(2)})
is just the flavor trace ${\rm tr}({\cal Q}^2\lambda^3/\sqrt{2})$ 
where ${\cal Q}$ is the matrix of the fractional quark charges.

In Ref. \cite{KeKl1} (where only the BC vertex was employed), the 
transition form factor $T_{\pi^0}(-Q^2,0)$ was numerically evaluated 
(for $0 < Q^2 < 2.8$ GeV$^2$ only) employing the soft and chiral limit. 
That is, in Ref. \cite{KeKl1} we approximated (for the pion only!) the 
BS vertex by its leading, ${\cal O}(p^0)$ piece: $\Gamma_{\pi^0}(q,p)
\approx \Gamma_{\pi^0}(q,0) = \gamma_5 \, \lambda^3 \, B(q^2)/f_\pi$.
For the sake of comparison, we again give $T_{\pi^0}(-Q^2,0)$ evaluated 
in that approximation (denoted by the dash-dotted line in Fig. 2),
but now up to $Q^2 \approx 8$ GeV$^2$. However, in the present work we go 
beyond this approximation, using our complete solution for the BS vertex 
$\Gamma_{\pi^0}(q,p)$, {\it viz.} the BS amplitude $\chi_{\pi^0}(q,p)$, 
given by the decomposition into 4 scalar functions multiplying independent 
spinor structures. The approximation we still keep, is discarding the 
second and higher derivatives in the momentum expansions.  

\noindent {\bf 4.}
$T_{\pi^0}(-Q^2,0)$ obtained in this way, and using the BC vertices, is 
depicted (after multiplication by $Q^2$) in Fig. 2 by the solid line. 
It tends to be on the high side 
of the data, but the general agreement of our $T_{\pi^0}(-Q^2,0)$ with 
the experimental points is rather good, except for the intermediate 
momenta around the interval from 1 to 2 GeV$^2$ interval. 
The dashed line denotes $Q^2 T_{\pi^0}(-Q^2,0)$ obtained in the same way,
but employing the mCP vertices.
For both curves, the agreement with experiment would be improved by
lowering them somewhat (at least in the momentum region $Q^2\lsim 4$
GeV$^2$), 
which could be achieved by modifying the model \cite{jain93b} and/or its 
parameters so that such a new solution for $A(q^2)$ is somewhat lowered 
towards its asymptotic value $A(q^2\to -\infty) \to 1$. (Of course, in order
to be significant, this must not be a specialized refitting aimed {\it only}
at $A(q^2)$. Lowering of $A(q^2)$ must be a result of a broad fit 
to many meson properties, comparable to the original fit \cite{jain93b}.
This, however, is beyond the scope of this letter.)

While the empirically successful anomaly result 
(\ref{AnomAmpl}) at $Q^2=0$ is model--independent in any 
consistently coupled SD-BS approach (being the consequence
only of the correct chiral behavior due to the incorporation of
D$\chi$SB), the transition form factor for $Q^2\neq 0$ of 
course depends not only on the WTI-preserving $qq\gamma$-vertex 
Ansatz, but also on the chosen bound-state model. However, it must 
be stressed that we we did not do any parameter fitting whatsoever, 
as we have used the parameters obtained from Jain and Munczek's 
\cite{jain93b} broad fit to the meson spectrum (except the
$\eta$--$\eta^\prime$ complex) and the pseudoscalar decay constants. 
Our model choice \cite{jain93b} has subsequently given us predictions 
for $\eta_c,\eta_b\to\gamma\gamma$ \cite{KeKl1}, the description of
$\eta$--$\eta^\prime$ complex and $\eta,\eta^\prime\to\gamma\gamma$ 
\cite{KlKe2}, and now, with the same set of parameters, 
the transition form factors which are empirically successful for 
the presently accessible high values of $Q^2$, and at the same time
in agreement (in the chiral limit even exactly and analytically 
\cite{bando94,Roberts,Frank+al,KeKl1,KlKe2}) with 
the anomalous amplitude (\ref{AnomAmpl}) at $Q^2=0$. 

Our transition form factors thus also agree rather well with other 
successful theoretical approaches. The vector--meson dominance (VMD) 
model (which is still the most successful one from the purely phenomenological 
point of view \cite{H+Kinoshita98})
and the QCD sum rule approach \cite{Radyushkin+Rusk3} give the 
transition form factor which is some 10\% below our ``BC" $T_{\pi^0}(-Q^2,0)$ 
for large values of $Q^2$. Therefore, in the large $Q^2$ region, our 
``BC" values are halfway between the uppermost line in Fig. 2 denoting 
the asymptotic pQCD \cite{BrodskyLepage} version (${\cal J} = 2$) 
of the result (\ref{largeQ2}), and VMD as well as the QCD sum rule results
\cite{Radyushkin+Rusk3} of Radyushkin and Ruskov [amounting to Eq.
(\ref{largeQ2}) with ${\cal J}\approx 1.6$].
By the highest presently accessible momenta, our ``mCP" version of 
$T_{\pi^0}(-Q^2,0)$ has crossed to the low side of both other theoretical 
predictions and the data, but is still within the error bars. 

The large-$Q^2$ leading power-law behavior (\ref{largeQ2}) was 
first derived from the parton picture in the 
infinite momentum frame -- {\it e.g.}, see Ref. \cite{BrodskyLepage}.
In this and other similar pQCD approaches, 
the precise value of the coefficient of the leading $1/Q^2$ term
depends on the pion distribution amplitude 
$\varphi_\pi(x)$ which should contain the necessary nonperturbative
information about the probability that a partonic quark carries the
fraction $x$ of the total longitudinal momentum. 
{\it E.g.}, the well-known example of a ``broad" distribution
$\varphi^{CZ}_\pi(x)= f_\pi 5 \sqrt{3} x(1-x)(1-2x)^2$ 
(proposed by Chernyak and Zhitnitsky \cite{ChernyakZhit} motivated 
by sum-rule considerations) leads to ${\cal J}=10/3$, but it
is too large in the light of the latest CLEO data \cite{gronberg98}.
In contrast, the asymptotic
$\varphi^{A}_\pi(x)= f_\pi \sqrt{3} x(1-x)$ favored by 
Lepage and Brodsky \cite{BrodskyLepage} yields ${\cal J}=2$, resulting in 
the line touching the error bars of the presently highest $Q^2$ data 
from above in Fig. 2. 
In fact, in the {\it strict} $\ln Q^2\to\infty$ limit, every 
distribution amplitude must evolve into the asymptotic one,
$\varphi_\pi(x) \to \varphi^{A}_\pi(x)$, if the effects of the 
pQCD evolution are taken into account. 
However, even at $Q^2$-values considerably larger than the presently
accessible ones, other effects may still be more important than the
effects of the pQCD evolution. This is the reason why other approaches
and other forms of $\varphi_\pi(x)$ should be considered even when they
do not incorporate the pQCD evolution.
The form 
\begin{equation}
\varphi_\pi(x) = \frac{f_\pi}{2 \sqrt{3}} 
              \frac{\Gamma(2\zeta+2)}{[\Gamma(\zeta+1)]^2}
               x^\zeta (1-x)^\zeta \, , 
           \quad \zeta > 0 \, ,
\label{genForm}
\end{equation}
is suitable for representing various distribution amplitudes because 
it is relatively general \cite{BrodskyLepage}: 
$\zeta > 1$ yields the (empirically favored) distributions that are 
``peaked" or ``narrowed" with respect to the asymptotic one ($\zeta = 1$), 
whereas $\zeta < 1$ gives the ``broadened" distributions which, however, 
now seem to be ruled out for the same reason as $\varphi^{CZ}_\pi(x)$ 
quoted above, since ${\cal J} > 2$ is ruled out empirically by CLEO 
\cite{gronberg98}. Namely, it is easy to see that Eq.~(\ref{genForm}) 
implies $\zeta = 2/(3 {\cal J} - 4)$. 

Indeed, at the highest presently accessible momenta, the asymptotic 
prediction (${\cal J} = 2$) is
lowered by some 20\% by the lowest order QCD radiative corrections
\cite{Brodsky+al98}, amounting to ${\cal J}\approx 1.62$, which fits the 
CLEO data well. Of course, these ${\cal O}(\alpha_s)$ corrections mean 
that $Q^2 T_{\pi^0}(-Q^2,0)$ is not strictly constant, but according to 
Ref. \cite{Brodsky+al98} it rises towards $2 f_\pi$ so slowly 
(by just 4\% from $Q^2=9$ GeV$^2$ to $Q^2=36$ GeV$^2$) 
that we can take it constant in practice. 
The situation is similar with the sum-rule approach of Radyushkin and 
Ruskov \cite{Radyushkin+Rusk3}. Their $Q^2 T_{\pi^0}(-Q^2,0)$ starts
actually {\it falling} after $Q^2\sim 7$ GeV$^2$, but so slowly that 
Eq. (\ref{largeQ2}) with the constant ${\cal J}\approx 1.6$ represents
it accurately at the presently accessible values of $Q^2$. 
This corresponds to a rather narrow distribution (\ref{genForm})
with $\zeta = 2.5$. 

The same type of the leading large-$Q^2$ behavior as in Eq. (\ref{largeQ2}),
was obtained by Manohar \cite{manohar90} using the operator product expansion 
(OPE). 
According to his OPE calculation, the coefficient in Eq. (\ref{largeQ2}) 
giving the leading term is ${\cal J}=4/3$, which is below 
our large-$Q^2$ $T_{\pi^0}(-Q^2,0)$ by 
$\sim 20\%$ when we use the BC vertex, but -- as we will see below --
exactly coincides with the $Q^2\to\infty$ limit obtained  
using the mCP and CP vertices. 
The coefficient ${\cal J}=4/3$ is the lowest one still consistent with the 
form (\ref{genForm}) because it corresponds to $\zeta = \infty$. The pion 
distribution amplitude (\ref{genForm}) then becomes infinitely peaked 
delta function: $\varphi_\pi(x)=({f_\pi}/{2 \sqrt{3}})\, \delta(x - 1/2)$.

For $Q^2 > 4$ GeV$^2$, our ``BC" $T_{\pi^0}(-Q^2,0)$ also 
behaves in excellent approximation as Eq.~(\ref{largeQ2}), with 
${\cal J} \approx 1.78$. This ${\cal J}$ would, in the pQCD 
factorization approach, 
correspond to $\varphi_\pi(x)$ (\ref{genForm}) with $\zeta=1.5$. 
On the other hand, our ``mCP" $T_{\pi^0}(-Q^2,0)$ falls off
faster than $1/Q^2$ for even the largest of the $Q^2$ values 
depicted in Fig. 2. However, it does not fall off much faster,
as our ``mCP" $Q^2 T_{\pi^0}(-Q^2,0)$ at $Q^2=18$ GeV$^2$ 
is only 6\%, and at the huge $Q^2=36$ GeV$^2$ is only
10 \% smaller than at $Q^2=9$ GeV$^2$ ({\it i.e.}, at the highest
presently accessible momenta).
Moreover, we can show analytically that, generally,  
the $1/Q^2$-behavior of Eq. (\ref{largeQ2}) must at some point 
be reached in our approach, although Fig. 2 shows that for the 
mCP $qq\gamma$ vertices this can happen only at significantly 
higher $Q^2$ than it happens for the BC $qq\gamma$ vertices.

\noindent {\bf 5.} It is very pleasing that in the present approach
(the coupled SD-BS approach in conjunction with GIA),
besides numerical results, one can get also some analytical insights 
in the regime of asymptotically large $Q^2$. This way we can make more 
illuminative comparisons with the asymptotic behaviors predicted by 
the various approaches -- notably pQCD on the light cone 
\cite{BrodskyLepage,ChernyakZhit} and OPE \cite{manohar90}.

Since the pion is light, $k\cdot k^\prime \approx Q^2/2$
for large negative $k^{2}=-Q^2$ and ${k^\prime}^2=0$.
Taking into account the behavior of the propagator functions 
$A(q^2)$ and $B(q^2)$ ({\it e.g.}, see\cite{KeBiKl98})
we can then in Eq. (\ref{Tmunu(2)}) approximate those quark propagators 
that depend on the photon momenta $k$ and $k^\prime$, by their  
asymptotic forms:
\begin{equation}
S(q-\frac{p}{2}+k) \approx S(k-\frac{p}{2}) \approx
-\frac{2}{Q^2}(\slashed{k}-\frac{1}{2}\slashp)~,
\label{asympS}
\end{equation}
and analogously for the propagator where $k$ is replaced by $k^\prime$.
Although the relative loop momentum $q$ can be large in the course of 
integration, its neglecting is justified because 
the BS amplitude $\chi(q,p)$ decays quickly and thus strongly 
damps the integrand for the large $q'$s.

For the moment let us make an additional approximation by replacing 
the dressed electromagnetic vertices in Eq.~(\ref{Tmunu(2)}) by the 
bare, free ones: $\Gamma^\mu(q,q^\prime) \to \gamma^\mu$. 
The tensor amplitude 
(\ref{Tmunu(2)}) then becomes
\begin{equation}
T_{\pi^0}^{\mu\nu}(k,k^\prime) = 
\frac{\sqrt{2}}{Q^2} \int\frac{d^4q}{(2\pi)^4}
\tr \left( \chi(q,p) 
\left[ (k-\frac{p}{2})_\lambda \gamma^\mu\gamma^\lambda\gamma^\nu
+ (k^\prime-\frac{p}{2})_\lambda \gamma^\nu\gamma^\lambda\gamma^\mu
\right] \right) \, ,
\label{asyStep1}
\end{equation}
which is readily rewritten as
\begin{equation}
T_{\pi^0}^{\mu\nu}(k,k^\prime)
=
- \frac{i\sqrt{2}}{Q^2}\,
\varepsilon^{\mu\lambda\nu\sigma}
(k-k^\prime)_\lambda
\int\frac{d^4q}{(2\pi)^4}
\tr [\chi(q,p)\gamma_\sigma\gamma_5]
= - \frac{2}{N_c \, Q^2}\,\varepsilon^{\mu\lambda\nu\sigma}
(k-k^\prime)_\lambda p_\sigma \, f_\pi~.
\label{TQ2:T_munu}
\end{equation}
The second equality holds since 
in the Mandelstam formalism, the integral in Eq.~(\ref{TQ2:T_munu}) is 
equal to $-i p_\sigma f_\pi \, \sqrt{2}/N_c$ 
-- {\it e.g.}, see \cite{jain91+munczek92,KlKe2}. 
Using the definition 
of the scalar amplitude $T_{\pi^0}(k^2,{k^\prime}^2)$, 
we finally recover Eq. (\ref{largeQ2}) with 
the coefficient having the special value ${\cal J}=4/3$.
We have thus found that the asymptotic behavior predicted by the 
present approach {\it with bare $qq\gamma$ vertices} 
(and, as seen below, also with dressed mCP and CP vertices), 
is in exact agreement with the leading term predicted by OPE \cite{manohar90}.

The asymptotic behavior $\propto 1/Q^2$
obtained in the present approach is especially 
satisfying when compared with that resulting from the calculation of the 
triangular quark loop carried out in the simple constituent quark model 
(with the constant light--quark mass parameter $m_u$), where 
$T_{\pi^0}(-Q^2,0) \propto (m_u^2/Q^2)\ln^2(Q^2/m_u^2)$
as $Q^2\to\infty$, which overshoots the data considerably
\cite{H+Kinoshita98} because of the additional $\ln^2(Q^2)$-dependence.

As seen from Fig. 2, the asymptotic behavior
$Q^2 T_{\pi^0}(-Q^2,0) = (4/3) f_\pi $ 
barely touches the experimental error bars 
from below. However, Manohar \cite{manohar90} pointed out that his
OPE approach also indicates the existence of potentially large 
corrections to his leading term. 
(Note that he also 
pointed out that, for the same reason, significant corrections 
should similarly affect the light-cone prediction (${\cal J}=2$) of 
Lepage and Brodsky \cite{BrodskyLepage}.) It is easy to see that 
in our case when the BC vertex is used, the main origin of the 
corrections raising ${\cal J}$ from ${\cal J}=4/3$ resulting from the usage 
of the bare vertices, to ${\cal J}\approx 1.78$ of our full numerical 
calculation, is the usage of these dressed $qq\gamma$ vertices. It is 
nevertheless instructive to formulate another successive approximation 
to illustrate the gradual transition to the $T_{\pi^0}(-Q^2,0)$ 
resulting from the full calculation. The idea is that at high
$Q^2$, only the $H_+$-term contributes significantly in the 
BC-vertex (\ref{BC-vertex}), as the $H_-$-terms are suppressed by 
their $Q^2$-dominated denominators in the both BC-vertices: 
$p_1^2 - p_2^2 = \pm (q \mp p/2)^2 \mp (k + q - p/2)^2 \approx \pm Q^2/2$.
The detailed derivation shows that, as $Q^2$ grows, this                    
indeed happens, and only
 $ \frac{1}{2}\gamma_\mu [ A(p_1^2) + A(p_2^2)]$
contributes significantly to both of the BC vertices.
Since $p_1^2$ in one of the vertices, and $p_2^2$ in the other,
also contain $Q^2/2$, the BC vertices reduce to 
$\gamma_\mu [1+A([q\pm p/2]^2)]/2$ since $A(-Q^2/2)\to 1$
as $Q^2\to \infty$. The adequacy of the chiral limit approximation
$p^2=M_\pi^2=0$ (and, in the case of our particular model choice, 
of neglecting the derivatives of $A$ in the expansion of 
$A([q\pm p/2]^2)$  around  $q^2$ since our solution  for  $A(q^2)$
is a smooth, slowly varying function \cite{KeBiKl98}) 
leads to the asymptotic expression
\begin{equation}
T_{\pi^0}(-Q^2,0) \approx \frac{4}{3} \frac{{\widetilde f}_\pi}{Q^2}
\label{ampWtilde}
\end{equation}
where ${\widetilde f}_\pi$ is given by the same Mandelstam-formalism 
expression as $f_\pi$, except that the integrand is modified by the 
factor $[1+A(q^2)]^2/4$. A further approximation would be to substitute
$A(q^2)$ by some average value, say $A({\widetilde q}^2)$ where
${\widetilde q}^2$ is a value from the $-q^2$-interval $[0.5, 2]$
GeV$^2$
of the momenta that have the most influence on the integral that gives
$f_\pi$. The relation 
${\widetilde f}_\pi \equiv f_\pi [1+A({\widetilde q}^2)]^2/4$
can serve as a definition of ${\widetilde q}^2$. The asymptotic 
value can thus -- in the case of the BC vertex (\ref{BC-vertex}) 
-- be expressed as Eq. (\ref{largeQ2}) with 
${\cal J} = [1+A({\widetilde q}^2)]^2/3$.
In the case of our solutions, the slow and moderate variation 
of the shape of $A(q^2)$ (see Fig. 3 in Ref. \cite{KeBiKl98}) permits 
the approximation $A({\widetilde q}^2) \approx (A_{min} + A_{max})/2
= (1 + A_{max})/2$. In our special case this can be further approximated
by $A(0)$, leading to the estimate $A({\widetilde q}^2)\sim A(0)=1.25$
\cite{KeBiKl98}.
This approximation would imply ${\cal J} \approx 1.69$,
whereas the more accurate Eq. (\ref{ampWtilde}) through
${\widetilde f}_\pi \approx 1.334 f_\pi$ leads to 
Eq. (\ref{largeQ2}) with ${\cal J}\approx 1.779$, 
which is in fact numerically indistinguishable from the asymptotics 
of our full calculation and the empirical fit. 

However, all this happens for the BC vertex, where the ``soft" leg 
adjacent to the pion BS-amplitude always contributes $A(q^2)$, but 
not for the mCP vertex, nor for the CP one. Since 
they tend to the bare vertex as soon as 
the high momentum flows through one of its legs, $f_\pi$ does 
{\it not} get replaced by ${\widetilde f}_\pi$, and the ``bare" 
result, Eq. (\ref{largeQ2}) with ${\cal J}=4/3$, 
continues to hold for $Q^2\to\infty$
when the mCP or CP vertex is used. Of course, the usage of 
the mCP vertex or the CP vertex
instead of $\gamma^\mu$ causes considerable differences for the
finite $Q^2$.

\noindent {\bf 6.}
The most important result of the present paper 
is that we have demonstrated that 
the modern version of the constituent quark model which is given by the 
coupled SD-BS approach, provides from $Q^2=0$ to $Q^2\to \infty$ the 
description for $\gamma\gamma^\star\to\pi^0$ which is -- 
independently of model details -- consistent with the Abelian ABJ 
anomaly and with the QCD predictions (\ref{largeQ2}) for the leading
large-$Q^2$ behavior. 
Since $f_\pi$ is a calculable quantity in the SD-BS approach,
the model dependence is present in the (successfully reproduced
\cite{jain93b,KeKl1,KlKe2,KeBiKl98}) {\it value} of $f_\pi$, but 
our derivation of the asymptotic {\it forms} (\ref{largeQ2}) and
(\ref{ampWtilde}) is model-independent. Of special importance is also
that this derivation [Eqs. (\ref{asyStep1}), (\ref{TQ2:T_munu})] of 
the large-$Q^2$ behavior applies to both Minkowski and Euclidean 
space. The same holds for the considerations involving ${\widetilde f}_\pi$
and the $[1+A({\widetilde q}^2)]^2/4$ correction factor, which have
enabled us to make smooth and accurate contact between our large-$Q^2$
numerical results calculated with the Euclidean solutions of the 
chosen model \cite{jain93b}, and analytical results on the large-$Q^2$
asymptotics. The agreement between the analytical and numerical results 
enhances our confidence in the accuracy of our numerical methods and
procedures (employing the Euclidean bound-state solutions) in this and
earlier papers \cite{KeKl1,KlKe2,KeBiKl98}.

To mention one possible application, the relevance of the present 
work for the light-by-light scattering contribution to the muon
$g-2$ is obvious. Nevertheless, the relevance of its extension to
the case when both photons are off-shell, {\it i.e.}, to the 
$\pi^0\gamma^\star \gamma^\star$  form factor, is even larger 
(for $g-2$), since there are no experimental data on 
$T_{\pi^0}(-Q^2, -{Q^{\prime}}^2)$ at present. In the light of its
consistency with OPE and axial anomaly, but also the (at least)
qualitative consistency with the QCD sum rules, light-cone pQCD 
and VMD dominance, we propose that the present approach be used 
in a re-calculation of some $\pi^0\gamma^\star\gamma^\star$-dependent 
predictions of Ref. \cite{H+Kinoshita98} instead of the double-VMD Ansatz
(or instead of the Extended Nambu-Jona-Lasinio model which does not give 
the correct asymptotic behavior even for the $\pi^0\gamma^\star\gamma$ 
vertex \cite{Bijnens+al9596}). The reason is that this VMD-motivated 
Ansatz for the $\pi^0\gamma^\star\gamma^\star$ form factor (Eqs. (3.5) and
(4.1) in Ref. \cite{H+Kinoshita98}) disagrees with what it should be according 
to the simple and unambiguous extension of $T_{\pi^0}(-Q^2,0)$ presented in  
this work. In particular, the VMD Ansatz for $T_{\pi^0}(-Q^2,-{Q^\prime}^2)$
behaves as $1/(Q^2 Q^{\prime 2})$ for asymptotically large $Q^2,
{Q^{\prime}}^2$,
whereas it should be only as the inverse of the sum of these squared momenta. 
Our full treatment of the $\pi^0\gamma^\star\gamma^\star$ form factor will be 
given in another, more detailed paper \cite{KeKlPrep1}, but already here we 
can point out that it is almost trivial to generalize the derivation of the 
asymptotic behaviors (\ref{largeQ2}) and (\ref{ampWtilde})  
to the case when both photons are off-shell, $k^2 = - Q^2 << 0$ and 
$k^{\prime 2} = - Q^{\prime 2} \leq 0$. Not much changes, except that 
$k\cdot k^\prime \approx (Q^2+{Q^{\prime}}^2)/2$.
Our prediction with the bare (or mCP, or CP) $qq\gamma$-vertices
for this generalization of Eq. (\ref{largeQ2}) turns out to be 
\begin{equation}
T_{\pi^0}(-Q^2, -Q^{\prime 2}) = \frac{4}{3} \, 
				\frac{f_\pi}{Q^2 + {Q^{\prime}}^2} \, ,
				\label{baregasgas}
\end{equation}                                
which is in agreement with the leading term of the OPE result 
derived by Novikov {\it et al.} \cite{novikov+al84} for the special 
case $Q^2=Q^{\prime 2}$. The distribution-amplitude-dependence 
of the pQCD factorization approach cancels out for that symmetric case, 
so that $T_{\pi^0}(-Q^2, -Q^{\prime 2})$ in this approach ({\it e.g.},
see \cite{Kess+Ong93}), in the limit $Q^2={Q^{\prime}}^2\to\infty$, 
exactly agrees with both our Eq. (\ref{baregasgas}) and Ref. 
\cite{novikov+al84}.
The inclusion of the dressed, WTI-preserving vertices does not require any
modifications of the {\it asymptotic} forms in the case of $qq\gamma$
vertices (such as mCP and CP ones) which reduce to the bare one 
as soon as one of their quark momenta squared tends to infinity.
In the case of the BC vertex 
(\ref{BC-vertex}), the generalization of the better-approximated 
expression (\ref{ampWtilde}) requires nothing except the substitution 
$f_\pi \to {\widetilde f}_\pi$ in Eq. (\ref{baregasgas}). 
For our \cite{KeBiKl98} particular $A(q^2)$, ${\widetilde f}_\pi$
can, in a good approximation, be factored as 
explained above.

We note that the VMD-motivated Ansatz for the 
$P^0\gamma^\star\gamma^\star$ ($P^0=\pi^0,\eta,\eta^\prime$) 
transition form factor is also used as a theoretical input in 
the CLEO data analysis -- see Eq. (8) in Ref. \cite{gronberg98}. 
A re-analysis of the data in the spirit of the present insights 
from the SD-BS approach, is therefore desirable.  
At least, one should examine the effects on the data analysis
when one makes a change in Eq. (8) of Ref. \cite{gronberg98}
like $1/[(1+Q^2/\Lambda^2)(1+{Q^\prime}^2/\Lambda^2)] \to
1/(1+Q^2/\Lambda^2+{Q^\prime}^2/\Lambda^2)$. 

Actually, $\eta$ and $\eta^\prime$ appear not only in the CLEO 
data \cite{gronberg98}, but are also relevant for the $g-2$ 
analysis \cite{H+Kinoshita98} discussed above.
This underscores the importance of extending the present work  
to include the 
$\eta\gamma^\star\gamma$ and $\eta^\prime\gamma^\star\gamma$
transition form factors. While the full calculation thereof 
obviously must be relegated to another paper \cite{KeKlPrep1},
our predictions for the $Q^2\to\infty$ behavior of the SU(3)$_f$
states $\eta_8$ and $\eta_0$, and also $\eta_c$ and $\eta_b$,
 can be easily obtained by redoing
the derivation of 
the asymptotic expression so that the track of their flavor 
content is kept. But, one then realizes that the 
generalization to both photons being virtual can  
be easily done also for $\eta_8,\eta_0,\eta_c$ and $\eta_b$
in the way discussed two passages above for $\pi^0$.
Therefore, in the approximation of the bare $qq\gamma$ vertices,
in the asymptotic regime with one of the photon momenta large (say
$Q^2\to\infty$), and the other (say
${Q^\prime}^2$) anywhere between 0 and $\infty$,
the $\eta_8$ and $\eta_0$ transition form factors are
given by the expressions analogous to Eq. (\ref{baregasgas}), 
except that its factor $4 f_\pi/3 \equiv C_{\pi^0}$, appropriate 
for $P^0=\pi^0$, is replaced by the respective factors
$C_{\eta_8} = (4/9\sqrt{3})(5 f_\pi - 2 f_{s\bar s})$ and
$C_{\eta_0} = (4\sqrt{2}/9\sqrt{3})(5 f_\pi + f_{s\bar s})$.
The auxiliary quantity $f_{s\bar s}$ is the decay constant of the unphysical 
$s\bar s$ pseudoscalar bound state, evaluated in \cite{KlKe2} to be 
$f_{s\bar s}= 1.47 f_\pi$ for the chosen bound-state model \cite{jain93b},
yielding the model-dependent values $C_{\eta_8} = 0.53 \, f_\pi = 49.3$
MeV and $C_{\eta_0} = 2.35 \, f_\pi = 219$ MeV.

When the $\eta_8, \eta_0$ asymptotic expressions are
improved by using the dressed, WTI-preserving vertices
instead of the bare vertex $\gamma^\mu$, the asymptotic expressions
do not change at all in the case of the mCP and CP vertex, 
and in the case of 
the BC-vertex the only changes are again just the substitutions
$f_\pi\to {\widetilde f}_\pi$ and $f_{s\bar s}\to {\widetilde f}_{s\bar s}$.

Similarly, for $\eta_c$ and $\eta_b$, the analogous expressions 
when at least one of $Q^2$ or ${Q^\prime}^2$ is large, are obtained 
from Eq. (\ref{baregasgas}) by replacing $4 f_\pi/3 \equiv C_{\pi^0}$ 
with $C_{\eta_c} = (16\sqrt{2}/9) {f}_{\eta_c}$ and
$C_{\eta_b} = (4\sqrt{2}/9) {f}_{\eta_b}$, respectively.
(Of course, if ${Q^\prime}^2<M_{P^0}^2$, neglecting the meson mass 
$M_{P^0}$ is for ${Q^\prime}^2<M_{P^0}^2$
a rougher approximation than setting ${Q^\prime}^2=0$.
$M_{\eta_c}$ and especially $M_{\eta_b}$ are already so large 
that keeping $k\cdot k^\prime = (Q^2+{Q^{\prime}}^2+M_{P^0}^2)/2$
in Eq. (\ref{baregasgas}) can be important for momenta $Q^2$
accessible in practice.) 
In our chosen bound-state model \cite{jain93b}, we have obtained 
${f}_{\eta_c} = 213$ MeV and ${f}_{\eta_b} = 284$ MeV, leading to 
the model-dependent values $C_{\eta_c} = 536 \,\, {\mbox{\rm MeV}}$ 
and $C_{\eta_b} = 179 \,\, {\mbox{\rm MeV}}$.
This is of immediate importance for experiment, as the $\eta_c$ 
transition form factor $T_{\eta_c}(-Q^2,0)$ can be measured at 
L3 \cite{Aurenche+al9601317}.


\section*{Acknowledgments}
\noindent The authors acknowledge 
the support of the Croatian Ministry of Science and
Technology contracts 1--19--222 and 009802 and discussions with 
T. Feldmann and R. Jakob.


\newpage

\section*{Figure captions}

\begin{itemize}

\item[{\bf Fig.~1:}] The diagram for $P^0 \to\gamma\gamma$
                decays ($P^0 = \pi^0,\eta, \eta^\prime, \eta_c, \eta_b$).
                Within the scheme of generalized impulse approximation,
                the propagators and vertices are dressed.

 \item[{\bf Fig.~2:}] 
The comparison of our results for $Q^2 T_{\pi^0}(-Q^2,0)$, 
the pion transition form factor, with the CELLO (circles) and 
CLEO (triangles) data and with the Brodsky-Lepage interpolation 
formula (the dotted line, which also denotes the pQCD limiting value 
$2 f_\pi$). The dash-dotted line represents our $Q^2 T_{\pi^0}(-Q^2,0)$ 
evaluated as in our earlier paper \cite{KeKl1} (but now to higher squared 
momenta), {\it i.e.}, exclusively with the BC $qq\gamma$ vertices, and 
in the chiral and soft limit approximation {\it for the pion}
[$\Gamma_{\pi^0}(q,p) \approx \Gamma_{\pi^0}(q,0) =
\gamma_5 \lambda^3 B(q^2)/f_\pi$]. 
Our results for $Q^2 T_{\pi^0}(-Q^2,0)$ without that approximation
are depicted by the solid line for the case of the BC vertices, and 
by the dashed line for the mCP vertices. The one obtained 
with the BC vertex practically saturates beyond $Q^2 \sim 6$ GeV$^2$ 
at higher values [$Q^2 T_{\pi^0}(-Q^2,0)\approx 164$ MeV] than that 
obtained with the mCP vertices. This latter one gets lower 
beyond $Q^2 \sim 2.5$ GeV$^2$ and does not yet saturate at the 
presently accessible momenta although approaches asymptotically 
$Q^2 T_{\pi^0}(-Q^2,0) \to 4 f_\pi/3$ for much higher $Q^2$.
(This limit, which is also the leading term resulting from OPE
\cite{manohar90}, is denoted by the dashed straight line.)

\end{itemize}


\newpage

\vspace*{4cm}
\epsfxsize = 16 cm \epsfbox{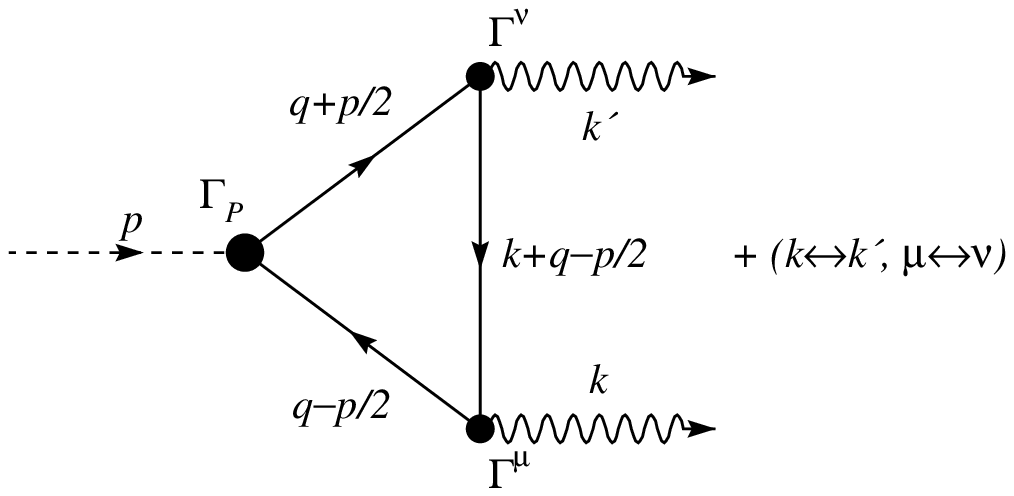}

\newpage

\vspace*{2cm}
\epsfxsize = 16 cm \epsfbox{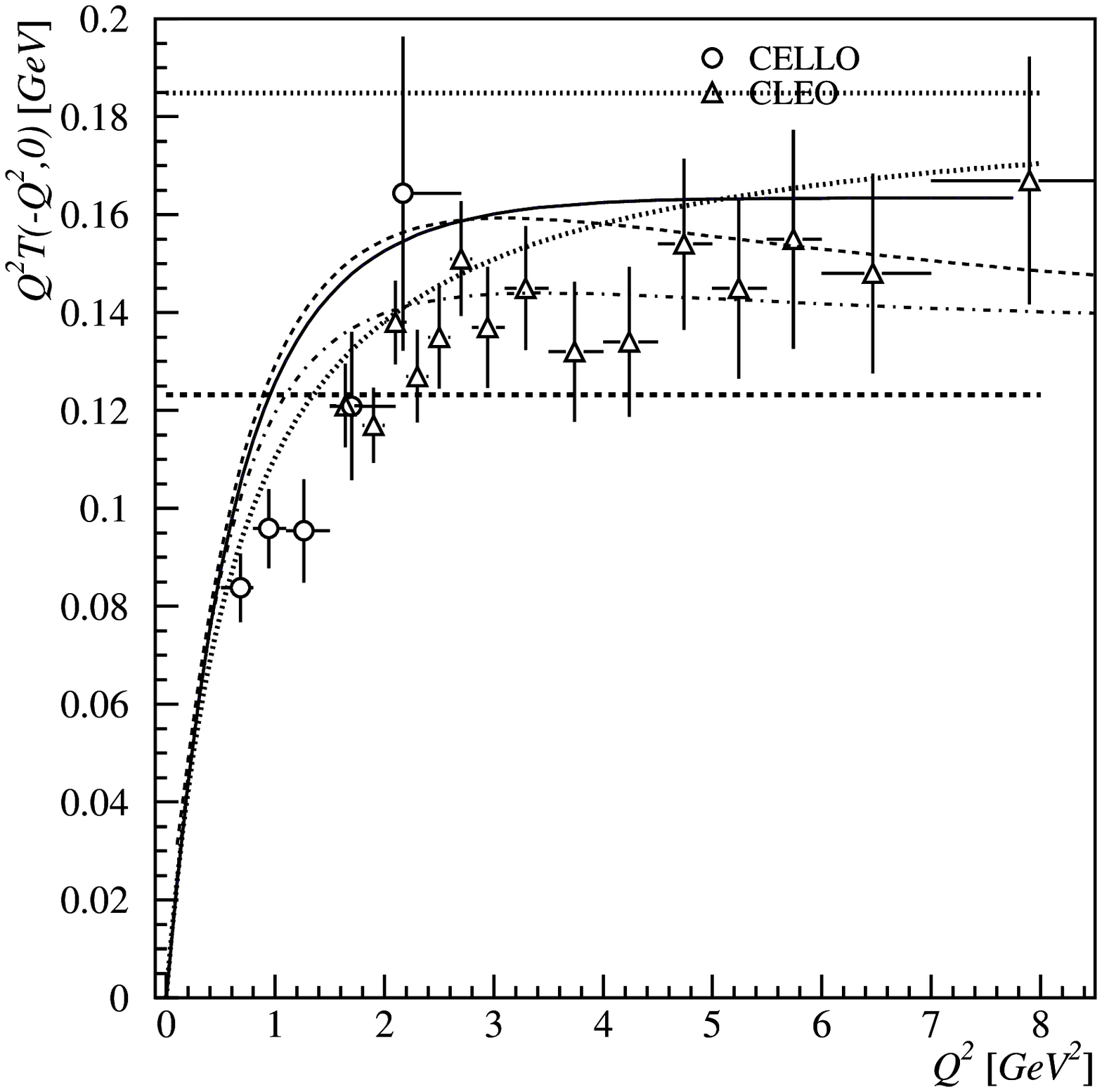}

\end{document}